# Lorentz invariance violation(LIV) in some basic phenomena in quantum physics

Z. Shafeei  and S. A. Alavi *

*Department of physics, Hakim Sabzevari university, P. O. Box 397, Sabzevar, Iran.*

*E-mail:* `s.alavi@hsu.ac.ir`

Lorentz symmetry is one of the cornerstone of both general relativity and the standard model of particle physics. We study the violation of Lorentz symmetry in some basic phenomena in atomic physics. Using the Green's function, and the source 4-current, the differential equation of 4-vector of electromagnetic potential is solved and the modified coulomb potential is obtained by some researchers. Using modified Coulomb potential, we find the corrections due to LIV on the spectrum of Hydrogen and Helium atoms. We also investigate the consequences of LIV on Stark, Zeeman and Spin orbit effects and obtain some upper bounds for the LIV coefficients.

## 1. Introduction

The Lagrangian density for the photon sector of the minimal SME is [1]:

$$L_{photon} = -\frac{1}{4}F_{\mu\nu}F^{\mu\nu} - \frac{1}{4}(K_F)_{k\lambda\mu\nu}F^{k\lambda}F^{\mu\nu} + \frac{1}{2}(K_{AF})^k \epsilon_{k\lambda\mu\nu}A^\lambda F^{\mu\nu} - j^\mu A_\mu \qquad (1)$$

In this equation, $J_\mu = (\rho, \vec{J})$ is the 4-vector current source that couples to the electromagnetic 4-potential $A_\mu$, and $F_{\mu\nu} = \partial_\mu A_\nu - \partial_\nu A\mu$ is the electromagnetic field strength. The coefficients $(K_F)_{k\lambda\mu\nu}$ and $(K_{AF})_k$ are assumed constant and control the CPT and Lorentz violation. To obtain a general solution for the potentials $\Phi$ and $\vec{A}$ we introduce Green functions $G_{\mu\lambda}(\vec{X},\vec{X}')$ so that:

$$A_\lambda = \int d^3X\, G_{\mu\lambda}(\vec{X},\vec{X}')j^\mu(X') \qquad (2)$$

Where for a point charge [1]:

$$G_{\mu\lambda}(\vec{X},\vec{X}') = \frac{\eta_{\mu\lambda} + (K_F)_{\mu j \lambda j}}{4\pi|\vec{X}-\vec{X}'|} - \frac{(K_F)_{\mu j \lambda k}(\vec{X}-\vec{X}')^j(\vec{X}-\vec{X}')^k}{4\pi|\vec{X}-\vec{X}'|^3} \qquad (3)$$

$\eta_{\mu\nu}$ is the metric of flat space. The source 4-current for a point charge is $\delta_0^\mu q \delta^3(\vec{X})$. By substituting this source and Eq.(3) in Eq.(2) and solving the integral equation for $\lambda=0$ we have [1]:

$$A_0(\vec{X}) = \frac{q}{4\pi|\vec{X}|}(1 - (K_F)_{0j0k}\hat{X}^j\hat{X}^k) \qquad (4)$$

In fact this is the modified Coulomb potential for a point charged particle. For $\lambda = j = 1,2,3$ we have [1]:

$$A_j(\vec{X}) = \frac{q}{4\pi|\vec{X}|}((K_F)_{0kjk} - (K_F)_{jk0l}\hat{X}^k\hat{X}^l) \qquad (5)$$

### 1.1 Effects of LIV on the spectrum of Hydrogen atom.

Comparing equation (4) with Coloumb potential one can find that the correction of LIV on Coloumb potential is given by $\frac{q}{4\pi|\vec{X}|}(K_F)_{0j0k}\hat{X}^j\hat{X}^k$. Since this correction if exists, should be very small compared to the Coulomb potential itself, so we can use time independent perturbation theory to calculate the effects of LIV on different phenomena in atomic physics. The shift of energy $\Delta E = \langle \phi_{nlm}|V|\phi_{nlm}\rangle$ on the nth hydrogen energy level is calculated as

$$\Delta E = \frac{q^2}{4\pi}(K)(\frac{1}{a_0 n^2}).$$

$a_0$ is the first Bohr radius. Since there has not been found any evidence for LIV in nature, so $\Delta E$ should be smaller than the accuracy of energy measurement $\Delta E \leq 10^{-12} ev$ [2], so we find the following upper bound on the LIV coefficient: $K \leq 2.8 \times 10^{-17}$.

For simplicity in our calculations, we have assumed the same values for all $(K_F)_{0j0k} = K$.

## 1.2 Consequences of LIV on Stark effect

The perturbed potential of Stark effect for a particle with charge q in an electric field E is $V = q\vec{E}.\vec{r}$. As mentioned before, LIV impose some corrections on Coulomb potential, so there is a correction on the electric field between electron and nucleus which can be named as permanent Stark effect because it is due to LIV not due to external electric field. The correction on electric field is given by:

$$E^j(\vec{X}) = \frac{q}{4\pi|\vec{X}|^2}(\hat{X}^j + 2(K_F)_{0j0k}\hat{X}^k - 3(K_F)_{0k0l}\hat{X}^l\hat{X}^k\hat{X}^j) \tag{6}$$

And finally after some calculations we arrived at the following expression for the corrections on the energy of nth energy level of hydrogen atom $\Delta E = -\frac{7q^2}{4\pi}K\frac{1}{a_0 n^2}$.

By the same arguments as previous section, we get the following upper bound for LIV coefficients: $k \leq 4.1 \times 10^{-18}$

## 1.3 LIV and spin-orbit interaction

The Hamiltonian of spin-orbit interaction is given by $H = \frac{-1}{2m_e^2}\frac{1}{r}\frac{d\phi(\vec{r})}{dr}\vec{S}.\vec{L}$. The Hamiltonian in the presence of LIV takes the following form:

$$H = \frac{1}{2m_e^2}\frac{1}{|\vec{x}|}\frac{1}{|\vec{x}|^2}\left(3 + 2(k_f)_{0j0k}\hat{x}^k\hat{x}^j - 9(K_F)_{0k0l}\hat{x}^l\hat{x}^k\right)\vec{s}.\vec{L} \tag{7}$$

Using this Hamiltonian and perturbation theory one can calculate the correction on the spin-orbit interaction energy:

$$\Delta E = \frac{1}{2m_e^2}(-7K) \times \left(\frac{\begin{cases}\frac{l}{2}\\-\frac{1}{2}(l+1)\end{cases}}{a_0^3 n^3 l(l+1)(l+1/2)}\right)$$

Considering $(n=2, l=1)$, the following upper bound on the coefficient k is obtained:

$$k \leq 8.7 \times 10^{-13}$$

## 1.4 Effects of LIV on the spectrum of Helium atom

The Hamiltonian of the Helium atom In the presence of LIV, takes the following form:

$$H = \frac{1}{2m}P_1^2 + \frac{1}{2m}P_2^2 - \frac{ze^2}{r_1}(1-(k_f)_{0j0k}\hat{x}^j\hat{x}^k) - \frac{ze^2}{r_2}(1-(k_f)_{0j0k}\hat{x}'^j\hat{x}'^k)$$

$$+ \frac{e^2}{|r_1-r_2|}(1-(k_f)_{0j0k}\hat{x}''^j\hat{x}''^k)$$

(8)

Using perturbation method, the corrections on the energy spectrum of Helium atom is obtained as follows $\Delta E = q^2 K \frac{3}{4a_0}$ .

Which leads to the following upper bound on the LIV parameter: $K \leq 3.8 \times 10^{-17}$

**Conclusions**

Using the modified Coloumb potential in the presence of LIV, we calculated the corrections due to LIV on the spectrum of hydrogen and helium atoms. We also calculated the LIV corrections on Stark and spin-orbit effects. Using the accuracy of energy measurment we obtained upper bounds on LIV coeficients for each aforementioned systems. The interesting point is that the upper bounds obtained from different systems are consistent.

| Quantum system or Quantum effect | The upper bound on LIV parameter |
|---|---|
| Hydrogen atom | $K \leq 2.8 \times 10^{-17}$ |
| Helium atom | $K \leq 3.8 \times 10^{-17}$ |
| Stark effect | $k \leq 4.1 \times 10^{-18}$ |
| spin-orbit interaction | $k \leq 8.7 \times 10^{-13}$ |